\documentclass[preprint,pra,letterpaper,superscriptaddress,floatfix]{revtex4}
\usepackage{graphicx,psfrag,bbm,latexsym,color,dcolumn,bm,dsfont,bbm,color,mathrsfs,bbold,latexsym,amsmath,amsfonts,amssymb,epsfig}

\newcommand{\beq}{\begin{equation}}
\newcommand{\eeq}{\end{equation}}

\newcommand{\beqa}{\begin{eqnarray}}
\newcommand{\eeqa}{\end{eqnarray}}

\begin{document}

\title{The Lattice $\beta$-function of Quantum Spin Chains}
\author{P.R. Crompton}
\affiliation{Institut f\"ur Theoretische Physik I., Universit\"at Hamburg, D-20355 Hamburg, Germany.}
\affiliation{Center for Theoretical Physics, Massachusetts Institute of Technology, Cambridge, MA  02139, USA.}
\vspace{0.2in}
\date{\today}

\begin{abstract}
{We derive the lattice $\beta$-function for quantum spin chains, suitable for relating finite temperature Monte Carlo data to the zero temperature fixed points of the continuum nonlinear sigma model. Our main result is that the asymptotic freedom of this lattice $\beta$-function is responsible for the nonintegrable singularity in $\theta$, that prevents analytic continuation between $\theta=0$ and $\theta=\pi$. } 
\end{abstract}

\maketitle

Haldane's conjecture for antiferromagnetic (AFM) spin chains predicts the existence of a nonvanishing gap for the groundstate of integer spin chains, and a vanishing gap for half-integer spin chain systems \cite{Haldane}. For a bond-alternating integer chain, for example, different topological realisations of the gapped groundstate are seen, separated by critical features driven by quantum fluctuations \cite{nakamura}. The low-energy effective model for the groundstate of these spin chain systems is the 2d O(3) model with a $\theta$-term. In the derivation of this effective action from the spin Hamiltonian in \cite{Haldane} the $\theta$-parameter is a function of 
both the lattice interaction couplings and the nonperturbative dynamics. At $\theta=0$ the results of the 2d O(3) nonlinear sigma model are in found to be in agreement with predictions for integer systems, through numerical analysis. However, if we vary the lattice interaction couplings to drive the quantum fluctuations it also drives nonlocal fluctuations of the vacuum, and the physics at $\theta =\pi$ can be difficult to investigate numerically because of the complex action problem. A similar difficulty is found in investigations of the 2d O(3) model in \cite{Chak}. Although quantum fluctuations are not relevant in this case, the experimentally relevant results derived for the scaling of the correlation length, are only defined through the cutoff scale for the nonlocal fluctuations. As such, the constants that appear in the nonlinear sigma model action (spin stiffness, spin wave velocity etc.) can 
only determined from phenomenological input, which serves to quantify the scale of the nonlocal fluctuations.
Perturbatively, it is possible to treat expansions around $\theta =\pi$ via deformations of the Sine-Gordon model with an irrelevant nonlinear term \cite{fat} to quantify the renormalization scale of the quantum fluctuations. However, is is only recently that the effect of nonlocal fluctuations has been considered in addition to quantum fluctuations for quantum spin chains \cite{mus1}\cite{mus2}. This has lead to the introduction of a double Sine-Gordon model description of the quantum spin chains in the vicinity of $\theta =\pi$ which leads to a change of the perturbation. What we want to consider in this article is what happens if the couplings of the perturbative expansion in \cite{fat} become strong due to nonlocal fluctuations, is it then possible to define a nonperturbative renormalization program.

The nonlinear sigma model, as a continuum theory, is known to be integrable at only two points $\theta=0$ and $\theta=\pi$ \cite{poly}. The nonintegrable singularities found as a function of varying $\theta$ mean that the changes of the groundstate topology cannot properly be described analytically as a function of varying $\theta$. 
From this we might conclude that the behaviour at $\theta=\pi$ cannot be represented through the continuous variation of $\theta$ on the lattice of a finite nonperturbative system. However, this is to neglect the effective status of the mapping of the groundstate from the spin Hamiltonian into the nonlinear sigma model \cite{sigma}. Haldane's derivation in \cite{Haldane} involves defining the quantum fluctuations about only a generalised groundstate operator for quantum spin chains, and this uniquely defines the relevant modes for perturbative expansion. A close analogy for this treatment is chiral perturbation theory in the context of QCD, which is formulated in terms of an effective Lagrangian that describes an exchange theory of pions. Provided that these modes are the main exchange particle for the dynamics, the system is well-modelled. However, the difficulty is that the effective parameterisation of the modes may not be of direct relevance if we consider different dynamical regimes. Such as, for example, those that occur in the O(3) model when it exhibits separately gapful and gapless groundstates for quantum spin chains. The general problem is how we reconcile the definition of $\theta$ on the lattice with the continuum definition.

Two effective continuum theories can be perturbatively expanded in different modes. If there is an analytic connection between the two different effective continuum theories it can be specified in terms of the renormalized couplings. However, an effective perturbative theory, defined only in unrenormalized nonperturbative lattice couplings, does not have the same significance as a renormalized theory. It is only after the lattice system has been mapped to the continuum limit of vanishing lattice spacing that it becomes meaningful to make a direct comparison between the different effective parameterisations of the groundstate. The nonperturbative system will then have been explicitly regularized, and the same UV cutoff applied to both effective theories, meaning that the two different nonlocal parameterisations can be related through expansion about a common fixed point. Consequently it only becomes meaningful to discuss the analyticity of lattice interaction couplings once the lattice renormalization group $\beta$-function  has been defined \cite{W1}. The only way in general that we can treat nonlocal fluctuations in a lattice system is to go to the UV limit of the model, where we are then sure that a perturbative expansion of the corresponding action is weak, and the convergence of the expansion is unaffected by strong nonlocal fluctuations.

It has been known for a long time that the 2d O(3) model exhibits asymptotic freedom at strong coupling \cite{poly2}, and also that $\theta=\pi$ corresponds to a renormalization group fixed point of the model \cite{aff}. However, no connection has previously been established between asymptotic freedom and nonintegrability. In fact, since both integrable and nonintegrable systems display asymptotic freedom this would tend to imply that there is no connection. 
However, we are not considering a general origin for the nonintegrability of the action, but a very specific one.
What we can consider independently from other sources for quantum spin chains, are the nonintegrable singularities associated only with the imaginary-valued source term parameter $\theta$. In addition by treating such a system via a lattice formalism, rather than conformal field theory, we can attempt to quantify the effect of strong nonlocal fluctuations via defining a nonperturbative renormalization scheme. 

Such nonintegrable phase factor singularities are of central importance to defining gauge field theories, through the properties of the Bohm-Aharonov experiment \cite{yang}. As Haldane had considered \cite{hal}, the gapless groundstate of the quantum spin chain can be described through the properties of the Braid groups of the spins as they relate to these phase factors. This gives a description of the gapless state of quantum spin chains in terms of semions or anyons \cite{fqh}, both of which have fractional statistics. What we evaluate in this article is the renormalization group flow of these topological objects, now defined discretely on the lattice, and how the fixed points of this new lattice renormalization group, defined as a function of the phase factor, relate to the nonintegrable phase factor singularities. What we consider, from Haldane's AFM Hamiltonian derivation of the nonlinear sigma model \cite{Haldane}, is a renormalization group equation defined with $\theta$ as the local coupling constant \cite{VW}\cite{Azc}. This local coupling links the lattice sites when the spin operators are defined in the basis of the generalised source term for nonlocal fluctuations. For this treatment we use the nonlocal operator definitions given in \cite{pres}\cite{PRD}. Usually the global expectation of this phase factor is determined from coupling the system to a Chern-Simons term \cite{hal}, but in our case we will now consider how the nonintegrable phase factor singularities arise solely from consideration of the nonperturbative dynamics. It is our main result to show that if there are nonintegrable singularities in the phase factor $\theta$ the renormalization constants of the new $\beta$-function will be dependent on the nonperturbative expectation value of the local coupling $\theta$. The new lattice $\beta$-function will consequently be UV stable \cite{asymp}, and therefore asymptotically free. Essentially we argue that in order for the size of the nonlocal fluctuations to the expansion in \cite{fat} to be negligible, and the convergence ensured, we must consider expansion in the UV.

\section{analytic continuation in $\theta$}

There are two different perspectives to be reconciled to give a more complete understanding of the nonintegrable phase factor singularities that arise in quantum spin chains at $\theta=\pi$. On the one hand the renormalization group flow of the nonlinear sigma model is well understood. However, the derivation of this model from the spin Hamiltonian \cite{Haldane} parameterises the model in terms of nonlocal fluctuations about a generalised groundstate. It is not envisaged how excitations above the groundstate can be incorporated in this treatment, and different groundstates can only be realised by considering independent systems. These properties imply that nonintegrable singularities separate the gapless and gapful parameterisations of the groundstate, but they do not define a dynamical origin for these singularities. On the other hand any arbitrary values of the interaction couplings 
of the spin Hamiltonian can be inputted into a nonperturbative lattice system. A finite spectrum of excited states is realised, and there is no analytic origin for the nonintegrable singularities as a function of the interaction couplings. The nonintegrable phase factor singularities arise in a nonperturbative system solely from the dynamics. To identify critical points at certain values of the lattice interaction couplings \cite{nakamura} is to have defined the properties of the nonperturbative vacuum, and the renormalization properties of the nonlocal fluctuations. 

This problem of reconciling nonperturbative dynamics at $\theta=\pi$ with continuum theories has been extensively treated in lattice gauge theory \cite{VW}\cite{PRD}, where it is used discuss parity symmetry breaking. However, these ideas have not yet been applied to quantum spin chains. We believe we can now extend this picture because of the anyon description of the gapless groundstate, which is expressed in terms of the Bohm-Aharonov factor \cite{hal}. 
Where nonintegrable singularities now arise for these objects, in our discrete lattice treatment, the definition of the phase factor will be equivalent to that of a gauge field \cite{yang}. However, the nonintegrable singularities and gauge fields will only be strictly defined in this new picture as a limiting process of reaching the continuum theory. There have been several attempts to determine to what extent it is possible to analytically continue 
the lattice partition function in the phase factor, or equivalently, source term parameter $\theta$, but a difficulty arises in defining how the continuum limit is reached. The reason for this is that the lattice source term for quantum fluctuations is nonlocal, and does not correspond to a locally conserved current. Consequently the source term is dependent on the lattice volume, and this leads to a strong, uncontrolled volume dependence of $\theta$ \cite{Azc}. Our new work for both gauge field theories and quantum spin system treatments is to identify a new topological lattice $\beta$-function to express this volume-dependence in $\theta$. A simple illustration of our basic motivation is given in \cite{Azc}. The example is that of the effect of nonlocal fluctuations in the Ising model, and their renormalizability via a nonperturbative program. At an external field value of $h=0$ the low temperature phase of the system contains a singularity in the free energy. What we could consider is what would happen if instead of considering a small perturbation in real $h$ we instead considered a small perturbation in imaginary $h$. Via a Wick rotation a correspondence can be made with Euclidean-time. It is argued in \cite{Azc} that the saddlepoint equation that defines the free energy is oscillatory when the thermodynamic limit is taken. Thus the nonlocal dynamical fluctuations of the Ising model are essentially undefined, but we want to ask whether the same is necessarily true of quantum spin chains. 

\section{lattice $\beta$-function}
Our aim is to now link all three descriptions of the quantum spin chain; the generalised nonlocal lattice source term description of Haldane \cite{Haldane}, the anyon picture of noninteracting closed loops \cite{hal}\cite{an1}, and the discrete lattice formulation of nonperturbative dynamics \cite{VW}\cite{qmc}. To do this we write the lattice partition function in the basis of the generalised source term for quantum fluctuations \cite{pres},
\beq
\label{partitionz}
\mathcal{Z}_{L}(\beta) = \int {\rm{d}} [\theta] \,\,\,
{\rm{exp}}\!\left[\,\int_{0}^{\beta} B(\bm{n}_s)\,i\theta -V(\bm{n}_s)
\,\,ds \,\right]
\eeq
\beqa\label{Hilberta}
B(\bm{n}) & \equiv &
\sum_{(s,s')}^{L \otimes \Theta} \,\, \sum_{\sigma \in G}
\lambda_{ss'\sigma}(\bm{n})\frac
{\langle \bm{n}\oplus \bm{1}_{s\sigma} \oplus \bm{1}_{s'\sigma}| \,\theta\, \rangle}
{\langle \bm{n}| \,\theta \,\rangle}\, \\
\label{Hilbertb}
V(\bm{n}) & \equiv &
\sum_{(s,s')}^{L \otimes \Theta} \,\, \sum_{\sigma \in G}
\lambda'_{ss'\sigma}(\bm{n})\frac
{\langle \bm{n}\oplus \bm{1}_{s\sigma} \oplus \bm{1}_{\sigma s'}| \bm{n}
\rangle}
{\langle \bm{n}| \bm{n} \rangle}\,
\eeqa
Each of the spin operators in (2) and (3) are defined nonperturbatively through their matrix elements, from the 
generalised spin vectors that describe the state of the lattice $\{\bm{n}\}$. The lattice is defined over a larger dimensional space than the quantum spin chain $L\otimes\Theta$, where $L$ is the length of the spin chain and $\Theta$ is the multiplicity of a noncompact phase factor. The partition function definition in (1) is a direct representation of the 1+1 numerical continuous-time Quantum Monte Carlo method \cite{qmc}, defined in Haldane's nonlocal basis \cite{Haldane}. The first operator $B$ is defined in the basis of the generalised source term for quantum fluctuations, and is such is diagonal in the space of the phase factor $\theta$, (2). The second operator, $V$, contains all other interactions in the spin Hamiltonian that are nonlocal in this basis, (3). Both operators are projected out, or dimensionally reduced, in the action onto the basis of quantum fluctuations. This is defined by a path dependent integration over $s$, the local site index, defined upto the the lattice cutoff $\beta$, which is the inverse temperature for this treatment. The discrete spin indices of the operators are given by $\sigma$, defined as an element of the general discrete group $G$, where $B$ defines the $S_z$-component of the operators. 
The partition function in (1) is defined by a path integral for the dynamics by integrating over all noncompact phases, which is used to represent the $S_x$ and $S_y$ components of the spin operators. Consequently, the partition function is not restricted to any particular topological sector. 

The path integral step is the most important for relating this new formulation to anyons. Anyons have several unusual features, and key among these is the implication of the breaking of parity symmetry. This property basically arises for anyons because applying the Braid group to the phase factor leads to a path dependence on the manifold \cite{an1}.
A similar result naturally arises in the determination of the nonperturbative vacuum using numerical loop-cluster methods. A definite parity is locally realised on the 1+1 numerical system. This arises because the parity of loops is different whether the loops evolve over the $L$ or $\Theta$ boundaries. Basically from our definitions in (\ref{Hilberta}) and (\ref{Hilbertb}) the winding is defined mod($\pi$) for neighbouring spatial sites, since they relate antisymmetric neighbouring spins, but mod($2\pi$) for neighbouring time sites, since they relate antisymmetric neighbouring instantons. After dimensional reduction this symmetry is retained locally as a property of the winding associated with the loop-clusters starting from each site. The new nonperturbative description therefore essentially gives the same local picture, of parity as a consequence of path dependence on $L\otimes\Theta$ that was developed for the anyon and semion pictures, \cite{hal}\cite{an1}. Without introducing an explicit four-fold symmetric interaction term into the Hamiltonian \cite{Sachdev} there is still an implicit definite parity realised locally in the nonperturbative dynamics of the quantum spin chain, because the basis vectors are defined in nonlocal units. 

The finite lattice definition in (1) can also be viewed as the exact probabilistic definition of a sequence of Poisson processes that relate each lattice site, \cite{pres}. To make this association, the 1+1 system is assumed to be both holomorphic and meromorphic in $\theta$ and $\beta$. The singularities of the system, branch cuts and poles, therefore occur only at the vertices of the $L\otimes\Theta$ lattice. Both of the two lattice variables $\theta$ and, $\beta$ are otherwise continuous. Thus, a definition of lattice spacing $a$ can be given over both $L$ and $\Theta$ by viewing the source term parameter definition in (\ref{Hilberta}) as the action of parallel transport between sites \cite{Wils}, 

\beq
\theta \, a_{s} \equiv B(\bm{n}_s) -\langle B \rangle_{L,\Theta}\,, \quad
\beta  \, a_{s'} \equiv A(\bm{n}_s') - \langle A \rangle_{L,\Theta} 
\eeq
where $A$ is the operator equivalent of (2) defined in the local spatial basis of the spin Hamiltonian. By solving the partition function in (1) nonperturbatively $B$ is maximised in the continuum at some particular value of $\theta\equiv\theta^{\,0}$. (4) now gives a new way of defining this expectation for discrete lattice distributions, away from the limit of continuous $s$. This means that the discrete picture of lattice anyons, in which parity is realised locally and is now encoded in the local value of $a_{s}$ can now be related directly to the nonlocal picture of the global phase factor of the O(3) model, which is now encoded in $\theta^{\, 0}$. 

Following the usual lattice gauge field theory picture we can replace parallel transport with a first derivative, and from the Hausdorff formula we will then have three leading terms corresponding to a two-loop operator expansion of (1) to give the effective action,
\beqa
\label{S}
S & = & \int_{0}^{\beta} \!\! ds \, \partial_{s} B(\bm{n}_{s}) \, -\, V(\bm{n}_{s})\,
-\,\partial_{s} [V(\bm{n}_{s}) B(\bm{n}_{s})] \\
& =& \int_{0}^{\beta} \!\! \int_{-\pi}^{\pi} \!\! ds'' \quad [ 1 - \langle V(\bm{n_{s}})\rangle_{\Theta} ]\,\partial^{2}_{s''} B(\bm{n}) \nonumber \\ 
& - & \partial_{s''} \langle B(\bm{n}_{s})\rangle_{\Theta}\,\partial_{s''} 
\langle V(\bm{n}_{s})\rangle_{\Theta} -[ 1 - \langle B(\bm{n_{s}})\rangle_{\Theta}]\, \partial^{2}_{s''} V(\bm{n}) \\
\label{S1}
& =& \int_{0}^{\beta} \!\! \int_{-\pi}^{\pi} \!\! ds'' \,\, a_{s'}\,\partial^{2}_{s''} B(\bm{n}) - a_{s}\,\partial^{2}_{s''} V(\bm{n})   - 
a_{s}\wedge a_{s'}
\eeqa
We have performed two integration steps in (5) and (6). The first is the one which already appears in the exponent of the partition function in (1), defined over $\beta$ and the second arises from the path integration over all noncompact phases. This second integration is defined in the action over $s''$ which is the generalised site index on $L\otimes\Theta$. We must pass through the vertices in $\Theta$ to evaluate the integration of the action over the noncompact phase. This is well-defined in $\theta$ but not in $a_{s}$ which is defined locally in (4). Thus, the partition function is always analytic in all couplings, as we expect for a nonperturbative system, but the action is nonanalytic in the new lattice spacings, as we expect from the O(3) model. Although the action in (4) implies that the maxima of $B$ should only defined through a compact angle, the new partition function definition gives a different picture: one topological sector dominates through $\theta^{\, 0}$. This apparent discrepancy is resolved by identifying that the lattice partition function defined over all topological sectors in (1) has no physical meaning: only the continuum theory that it reaches. The only significant difference between our new effective lattice 
action and the irrelevant and double Sine-Gordon models in \cite{fat} and \cite{mus1}\cite{mus2} is that our action is defined in a nonlocal basis, but this means that we can very simply identify the renormalization group flow.

The new action can be evaluated analytically by redefining the contour for (7) to enclose the singularities associated with the vertices, using the residue theorem. To be clear, the expansion for the effective action is not in $\theta$ but is defined in the lattice spacing about the Gaussian behaviour assumed at $\theta^{\,0}$ whichever topological sector this corresponds to. As a consequence of making holomorphic and meromorphic assumptions for the partition function the background field defined on $L\otimes\Theta$ is trivial. The fixed point the action is expanded about in the UV through the lattice cutoffs $\theta$ and $\beta$ is therefore trivial, and the action asymptotically free. Thus, to assume that the effect of nonlocal fluctuations IR fixed point can be treated analytically, even for very strong nonlocal fluctuations, is to assume that a nonperturbative renormalization approach exists.

The renormalization group flow for (\ref{S1}) is simply given by \cite{BKT},
\beq
\label{RG}
\frac{da_{s}}{dl} = -\frac{1}{2} a_{s}^{2}\left(\frac{a_{s'}}{\pi}\right)^{2}, \quad
\frac{da_{s'}}{dl}=a_{s'}(2-2a_{s})
\eeq
where, $l={\rm{log}} L$. From the second of these two relations in (\ref{RG}) the renormalization group flow goes
to $a_{s'}=0$ for sufficiently large values of $a_{s}$, where $a_{s}$ is then effectively renormalized as $a_{s}\rightarrow a_{s}'$. We can therefore expect, in a realistic numerical simulation of (1), that a second order BKT effective renormalization region exists where the finite-temperature numerical system will map directly into the zero-temperature IR fixed point in (7). We would normally take it that the gap is renormalized in the vicinity of the IR fixed point, but this extends the result to show that regardless of the scale of the nonlocal fluctuations the singularities can be re-absorbed. Usually, in numerical studies that generate nonlocal fluctuations, it is assumed that the renormalization group flow implicitly converges to the IR fixed point by considering sufficiently large numerical systems such that Lorentz invariance is approximately realised. However, here we make a precise analytic connection suitable for evaluating the RG flow of modest finite size systems exactly. We have identified the modifications to the renormalization group flow by defining a nonperturbative renormalization group program, rather than by tackling the nonlocal fluctuations through computationally expensive step of simulating on large volumes in the UV limit.

\section{Mermin-Wagner theorem}
As a final remark we comment on the relationship between the fixed point of the O(3) model and the Mermin-Wagner theorem \cite{MW}. From the Mermin-Wagner theorem it is argued that we cannot have a finite-temperature quantum 
spin chain acquire long range order. Long-range order implies that the divergence of correlation length associated with the vanishing of the gap at $\theta=\pi$ would be prohibited at finite-temperatures. Seemingly the renormalization group flow in (8) contradicts this theorem because there is a finite $\beta$ mapping defined. 
However, in a later paper, Mermin argues correlations in the space of the multiplicity of vortices can be long-ranged at finite temperatures \cite{M}. The difference for these correlations is argued as being the range of the 
interaction: a finite-range interaction implies the absence of long range order, whereas an infinite-range interaction implies long-range ordering is possible. Crucially, the range of the interactions in our choice of nonlocal basis is effectively infinite, given in $L\otimes\Theta$. Correlations can in principle diverge faster than, $L$. There is therefore no contradiction with developing a new picture of defining the fixed point via finite temperature studies in quantum spin chains, simply again that it is only the continuum limit of the nonlocal basis where we can define the physical scale. 
\section{Summary}
Haldane has described how the groundstate of the system realised by an AFM chain Hamiltonian can be related to the O(3) model, which, as a continuum theory, is known to be integrable at only two points $\theta=0$ and $\theta=\pi$ \cite{poly}. Here we have inverted these arguments to investigate the O(3) model defined in terms of the unrenormalized lattice interaction couplings, which relate to the nonlocal fluctuations of the vacuum. We have derived a lattice $\beta$-function, given in terms of a nonlocal scale, via a nonperturbative renormalization program. We have identified that this lattice $\beta$-function must be asymptotically free in order to be convergent, and that it has an IR fixed point corresponding to the nonintegrable phase factor singularities of the 2d O(3) model. This gives a dynamical origin for the nonintegrability, and practically, a renormalization group flow that can be measured without having to perform numerical simulations on large volumes to investigate quantum criticality.

\end{document}